\documentclass[twocolumn,superscriptaddress,bibnotes,aps,floatfix]{revtex4}
\usepackage[hidelinks]{hyperref}

\usepackage{newtxtext}
\usepackage{newtxmath}
\usepackage[T1]{fontenc}

\usepackage{graphicx}
\usepackage{placeins}

\usepackage{amsmath}
\usepackage{amsfonts}
\usepackage{mathtools}
\usepackage{mathrsfs}
\usepackage{braket}

\DeclareMathOperator{\trx}{tr}
\DeclareMathOperator{\eig}{eig}

\def\Yset{\mathscr{C}}
\def\Rset{\mathscr{R}}

\def\abs#1{\lvert#1\rvert}

\def\emath{\mathrm{e}}

\begin{document}

\title{Witnesses of non-Gaussian features as lower bounds of stellar rank}

\author{Jan Provazn\'{i}k}
\affiliation{Department of Optics, Palack\'y University, 17. listopadu 1192/12, 77146 Olomouc, Czech Republic}

\author{\v{S}imon Br\"{a}uer}
\affiliation{Niels Bohr Institute, University of Copenhagen, Jagtvej 155 A, DK-2200 Copenhagen, Denmark}
\affiliation{Department of Optics, Palack\'y University, 17. listopadu 1192/12, 77146 Olomouc, Czech Republic}

\author{Vojt\v{e}ch Kala}
\affiliation{Centre for Quantum Information and Communication, \'{E}cole polytechnique de Bruxelles, CP 165, Universit\'{e} libre de Bruxelles, 1050 Brussels, Belgium}
\affiliation{Department of Optics, Palack\'y University, 17. listopadu 1192/12, 77146 Olomouc, Czech Republic}

\author{Jarom\'{i}r Fiur\'{a}\v{s}ek}
\affiliation{Department of Optics, Palack\'y University, 17. listopadu 1192/12, 77146 Olomouc, Czech Republic}

\author{Petr Marek}
\affiliation{Department of Optics, Palack\'y University, 17. listopadu 1192/12, 77146 Olomouc, Czech Republic}

\begin{abstract}
Quantum non-Gaussian states and operations serve as fundamental resources for universal quantum computation, error correction, and high-precision metrology, extending beyond the Gaussian limits. While the stellar rank provides a rigorous hierarchical measure of non-Gaussianity, it remains challenging to determine experimentally. Conversely, witnesses of non-Gaussian features, based on the expectation values and variances of measurable observables, offer an accessible method for certifying non-Gaussian behavior but lack a direct connection to stellar rank. In this work, we establish a quantitative connection between these witnesses and stellar rank, demonstrating that the former can provide certifiable lower bounds on stellar rank. We introduce normalized expectation value and variance-based quantifiers and show that these witnesses form a consistent hierarchy of thresholds corresponding to stellar rank. Our results bridge the gap between abstract hierarchical measures and experimentally accessible quantifiers, enabling scalable certification of non-Gaussian states.
\end{abstract}

\maketitle

%

\section{Introduction}

Bosonic modes of traveling light, described by continuous-variable (CV) quadrature operators, provide a robust, naturally scalable platform suitable for universal fault-tolerant quantum computation~\cite{gottesman2001,aghaeerad2025,larsen2025} and quantum metrology~\cite{podhora2022,giovannetti2004,pirandola2018,oh2020,hanamura2023}. Achieving a quantum advantage in these applications requires quantum non-Gaussian states~\cite{mari2012}, but alas, their reliable preparation remains one of the main experimental challenges of the platform~\cite{thekkadath2020,zhou2021,ourjoumtsev2007,miyata2016,aghaeerad2025,larsen2025}. While the experimental techniques for preparation of such states are limited to various configurations of linear optics, squeezed states, and photon number resolving detectors~\cite{dakna1999,fiurasek2005,filip2005,marek2009,marek2011,yukawa2013,yukawa2013b,miyata2016,marek2018,sakaguchi2023,jeong2015,su2019,eaton2019,konno2021,takase2021,endo2023,yoshikawa2018,tiedau2019,endo2025}, theoretical approaches to observing their non-Gaussianity are more varied, utilizing state decompositions~\cite{hahn2025}, negative areas in Wigner functions~\cite{genoni2013,hughes2014,straka2014,tan2020}, specific properties of photon number distributions~\cite{filip2011,albarelli2018,lachman2019,chabaud2020,chabaud2021,fiurasek2022}, and other numerous features discernible from reconstructed density matrices. 

Among the many existing approaches to certifying non-Gaussian features, stellar rank~\cite{chabaud2020,chabaud2021}, equivalent to genuine $n$-photonic quantum non-Gaussianity~\cite{lachman2019}, has emerged as a hierarchical and operationally motivated measure of non-Gaussianity, capturing the discrete structure of continuous-variable quantum states. Stellar rank of pure states is defined as the number of zeros of the Husimi Q-function. It can be interpreted operationally as the minimal number of photon additions required to generate the state from Gaussian resources~\cite{chabaud2020,chabaud2021,walschaers2021,fiurasek2022}. Its extension to mixed states is defined through a convex-roof construction, giving stellar rank a rigorous theoretical foundation but rendering it experimentally difficult to determine in practice.

Quantum non-Gaussian features can also be detected using witnesses based on statistical moments of quadrature operators, relying on expectation values or variances of experimentally accessible operators. Typical examples include Gottesman-Kitaev-Preskill (GKP) state squeezing~\cite{marek2024}, catability~\cite{brauer2025b}, generalized quadrature nullifiers~\cite{kala2025b}, and non-linear squeezing~\cite{konno2021,miyata2016,konno2024,brauer2021,brauer2025a,kala2025a,kala2025b}, with practical appeal of frequently requiring only a limited number of measurements, for example, of statistical moments of lower orders, often directly accessible on optical platforms without the need for full quantum state tomography. These criteria, introduced independently of one another, are motivated by state- or protocol-specific considerations and share the common goal of detecting statistical signatures that Gaussian states cannot reproduce, while only loosely connecting to the discrete hierarchy of stellar rank.

In this work, we demonstrate that witnesses of non-Gaussian features, based on statistical moments, yield rigorous lower bounds on stellar rank, and conversely, that the stellar rank imposes constraints on their values. We further show that both expectation-value and variance-based criteria form a monotonic hierarchy consistent with stellar-rank ordering. 

The rest of this paper is organized into a pair of primary sections. The first section introduces the theoretical methods, whereas the numerical results and findings are discussed in the latter part of the paper.

\section{Stellar rank and continuous witnesses}

A pure single-mode quantum state~$\ket{\psi_{m}}$ with stellar rank~$m$ can be expressed~\cite{fiurasek2022} as
\begin{equation}\label{psi}
    \ket{\psi_{m}} = \hat{G} \sum\limits_{n=0}^{m} c_{n} \ket{n}
    \,\text{,} 
\end{equation}
where~${\abs{c_{m}} > 0}$ and~$\hat{G}$ denotes an arbitrary single-mode Gaussian unitary operation, which can be decomposed~\cite{cariolaro2016} into a sequence~${\hat{G} \coloneqq \hat{F}(\vartheta) \hat{D}(z) \hat{S}(r) \hat{F}(\theta)}$ of elementary single-mode Gaussian operations
\begin{equation}
    \begin{gathered}
        \hat{F}(\vartheta) \coloneqq \exp(\imath \vartheta \hat{a}^{\dagger} \hat{a}) 
        \,\text{,} \quad
        \hat{D}(z) \coloneqq \exp(z \hat{a}^{\dagger} - \bar{z} \hat{a})
        \,\text{,} \\
        \text{and}\quad
        \hat{S}(r) \coloneqq \exp\left(\frac{r}{2}(\hat{a}^{2} - \hat{a}^{\dagger2})\right)
        \,\text{,}
    \end{gathered}
\end{equation}
where~$\hat{a}$ and~$\hat{a}^{\dagger}$ represent the bosonic annihilation and creation operators. The general Gaussian operation~$\hat{G}$ is specified by five real parameters in total. A pair of real parameters forms the complex displacement amplitude~$z$, the squeezing rate is given by~$r$, and finally, $\theta$ and~$\vartheta$ define the two phase shifts.

A mixed state has stellar rank \(m\) if it can be expressed as a statistical mixture of pure states with stellar rank at most~$m$, but not as a mixture of states with lower stellar rank~\cite{chabaud2020}. Stellar rank naturally forms a hierarchy of quantum states. Let~$\Yset_{m}$ denote the set of all states with stellar rank at most~$m$. These sets clearly satisfy~${\Yset_{m} \subset \Yset_{n}}$ for every~${m < n}$. The set of Gaussian states~$\Yset_{0}$ lies at the bottom of this hierarchy.

%

While stellar rank captures the discrete structure of a quantum state, witnesses based on the moments of directly measurable operators provide a complementary perspective. 

Let us first consider a witness based on the expectation value of an observable non-negative lower-bounded Hermitian operator~$\hat{W}$ and examine its minimal value achievable by quantum states with stellar rank~$m$
\begin{equation}\label{Wm}
    W_{m} 
    = \inf\limits_{\hat{\varrho}_{m} \in \Yset_{m}} 
        \trx ( \hat{\varrho}_{m} \hat{W} )
    \,\text{.}
\end{equation}
The practical evaluation of~\eqref{Wm} deserves further discussion. It depends linearly on~$\hat{\varrho}_{m}$, making it sufficient to consider only pure states~\eqref{psi} in the minimization
\begin{equation}
    W_{m}
    = \inf\limits_{\hat{G}, \ket{\psi_{m}} }
        \braket{\psi_{m} | \hat{G}^{\dagger} \hat{W} \hat{G} | \psi_{m}}
    \,\text{.}
\end{equation}
This approach can be further simplified~\cite{fiurasek2022}. Instead of minimizing over all the pure states~${\ket{\psi_{m}} \in \Yset_{m}}$, it is enough to find the least eigenvalue 
\begin{equation} \label{Wm-eig}
    W_{m} = \inf\limits_{\hat{G}} 
    \left[ 
        \min \eig \left(
            \hat{\Pi}_{m} \hat{G}^\dagger \hat{W} \hat{G} \hat{\Pi}_{m} \right) 
    \right]
\end{equation}
of the transformed operator~$\hat{W}$ projected onto the finite-dimensional Hilbert space spanned by the first~${(m+1)}$ Fock states using the projection operator
\begin{equation}
    \hat{\Pi}_m = \sum_{n=0}^m \ket{n}\!\bra{n}
    \,\text{.}
\end{equation}
The optimization in~\eqref{Wm-eig} is performed over all the possible single-mode Gaussian operations~$\hat{G}$, that is, at most over four real parameters instead of all five required to characterize the general Gaussian transformation. The right-most phase-shift~$\hat{F}(\theta)$ can be naturally absorbed into the eigen-decomposition as it maps the truncated Fock space onto itself, thus reducing the number of parameters by one. The required number of parameters could be further reduced due to symmetries within specific witness operators.

%

In general, linear witnesses can be improved by introducing additional non-linear terms~\cite{guhne2006}. Let us now consider a class of witnesses based on variance, specifically on non-linear squeezing, which effectively quantifies the suppression of fluctuations in a general Hermitian operator~$\hat{Q}$ beyond the limits attainable by some well-defined class~$\Yset$ of reference states~\cite{brauer2025a}. Non-linear squeezing offers experimental advantage as it can be observed directly in many relevant cases~\cite{miyata2016,konno2021,brauer2021,kala2022,konno2024,marek2024,brauer2025a,kala2025a}. It can be formally defined as
\begin{equation}\label{nls}
    \xi (\hat{\varrho}) = \frac{
        \langle \{\hat{Q} - \langle \hat{Q} \rangle_{\hat{\varrho}} \}^{2} \rangle_{\hat{\varrho}}
    }{
        \inf\limits_{\hat{\mu} \in \Yset} 
            \langle \{\hat{Q} - \langle \hat{Q} \rangle_{\hat{\mu}} \}^{2} \rangle_{\hat{\mu}}
    }
    \,\text{.}
\end{equation}
When the infimum is taken over the set of Gaussian states, the quantity serves as an indicator of non-Gaussianity~\cite{brauer2025a}. An arbitrary quantum state~$\hat{\varrho}$ is non-Gaussian if ${\xi (\hat{\varrho}) < 1}$.

The reference threshold values attainable by states with specific stellar rank~$m$ can be determined as
\begin{equation}
    V_{m} 
    = \inf\limits_{\hat{\varrho}_{m} \in \Yset_{m}} 
        \langle \{\hat{Q} - \langle \hat{Q} \rangle_{\hat{\varrho}_{m}} \}^{2} \rangle_{\hat{\varrho}_{m}} 
\end{equation}
where the variance depends quadratically on the quantum state. This can be resolved~\cite{cai1993,miyata2016} by introducing another real parameter~$\lambda$ into the optimization and considering the expectation value of the equivalent surrogate lower-bounded Hermitian operator~${(\hat{Q} - \lambda\hat{I})^2}$ instead, as detailed within the appendix,
\begin{equation}
    \begin{aligned}
        V_{m} 
        & = \inf\limits_{\lambda \in \Rset, \hat{\varrho}_{m} \in \Yset_{m}}
            \langle \{ \hat{Q} - \lambda\hat{I} \}^2 \rangle_{\hat{\varrho}_{m}} \\
        & = \inf\limits_{\lambda \in \Rset, \hat{\varrho}_{m} \in \Yset_{m}}
            \trx (
                \hat{\varrho}_{m} \{ \hat{Q} - \lambda\hat{I} \}^2 
            )
        \,\text{.}
    \end{aligned}
\end{equation}
This removes the quadratic dependency, making it possible to reuse the machinery from relation~\eqref{Wm-eig}. The reference value
\begin{equation} \label{Vm-eig}
    V_{m} = \inf\limits_{\lambda \in \Rset, \hat{G}} 
    \left[ 
        \min \eig \left(
            \hat{\Pi}_{m} \hat{G}^\dagger
            \{ \hat{Q} - \lambda\hat{I} \}^2 
            \hat{G} \hat{\Pi}_{m} 
        \right) 
    \right]
\end{equation}
is then obtained by optimizing over five real parameters at most.

The assumptions about the operators~$\hat{W}$ and~$\hat{Q}$ are rather loose. The operators induce witnesses of non-Gaussian features; it is therefore reasonable to further require that their global minima, understood as expectation values or variances, depending on the witness, are not attained by elements from the set of Gaussian states~$\Yset_{0}$. 

It is convenient to match the threshold~$V_{m}$ with the definition of non-linear squeezing~\eqref{nls} and normalize its value with respect to the same reference set~$\Yset$. Often, this is the set of Gaussian states~$\Yset_{0}$, and the thresholds are normalized with respect to the Gaussian limit~$V_0$, yielding the quantity~$\xi_{m} \coloneqq V_{m} V_{0}^{-1}$. In the same spirit, the witnesses based on expectation values can be also normalized, introducing a new quantity~${\zeta(\hat{\varrho}) \coloneqq \trx(\hat{\varrho} \hat{W}) W_{0}^{-1}}$ similar to non-linear squeezing. The corresponding thresholds~$\zeta_{m} \coloneqq W_{m} W_{0}^{-1}$ are also normalized with respect to the Gaussian limit~$W_{0}$.
 
Poincar\'{e} separation theorem gives a practical and straightforward interpretation of the thresholds. Consider an arbitrary quantum state~$\hat{\varrho}$. If~${\zeta(\hat{\varrho}) < W_m}$ holds, then~${\hat{\varrho} \notin \Yset_{m}}$. Similarly, if~${\xi(\hat{\varrho}) < \xi_{m}}$ holds, then~${\hat{\varrho} \notin \Yset_{m}}$. In both cases the state must necessarily have stellar rank at least~$(m+1)$.

%

\begin{table}[b]
    \def\arraystretch{1.5}
    \begin{tabular*}{\columnwidth}{@{\extracolsep{\fill}} llllr}
        \multicolumn{3}{l}{\textbf{Nullifier family}} & \textbf{Canonical prescription} & \\
        \hline\hline
        (a) & Cubic states & \cite{miyata2016,brauer2021,kala2022} & ${\hat{x} + \kappa \hat{p}^2}$ & \eqref{Vm-eig} \\
        (b) & GKP states & \cite{marek2024} & ${\sin^{2} (f_{x} \hat{x}) + \sin^{2}(f_{p} \hat{p})}$ & \eqref{Wm-eig} \\
        (c) & Cat states & \cite{brauer2025b} & ${( \hat{a}^{\dagger2} - \alpha^{*2})( \hat{a}^{2} - \alpha^{2}) + ( 1 \mp \hat{\Pi})}$ & \eqref{Wm-eig} \\
        (d) & Fock states & \cite{kala2025b,kalash2025,racz2025,hotter2025} & ${(\hat{n} - k)^2}$ & \eqref{Wm-eig} \\
        \hline
    \end{tabular*}
    \caption{Operator families and their representative canonical forms.}
    \label{refop}
\end{table}

\section{Discussion and numerical results}

\begin{table*}[t]
    \centering
    \renewcommand{\arraystretch}{1.2}
    \setlength{\tabcolsep}{6pt}
    \begin{tabular}{|c|c|c||c|c||c|c||c|c|}

      \multicolumn{1}{c}{}
    & \multicolumn{2}{c}{(a) $\kappa \equiv 1$} 
    & \multicolumn{2}{c}{(b) $f_{p} \equiv \sqrt{\pi} \equiv 2 f_{x}$} 
    & \multicolumn{2}{c}{(c) $\tau_{+},\, \alpha \equiv 2$} 
    & \multicolumn{2}{c}{(c) $\tau_{-},\, \alpha \equiv 2$} \\
    \cline{2-9}
    
    \multicolumn{1}{c|}{$\Yset_{m}$}
    & $V_{m}$ & $\xi_{m}$ & $W_{m}$ & $\zeta_{m}$ & $W_{m}$ & $\zeta_{m}$ & $W_{m}$ & $\zeta_{m}$ \\
    \hline 
    $m=0$  & 0.9449 & 1.0000 & 1.0000 & 1.0000 & 0.9997 & 1.0000 & 1.0003 & 1.0000 \\
    $m=1$  & 0.6774 & 0.7169 & 1.0000 & 1.0000 & 0.9997 & 1.0000 & 1.0003 & 1.0000 \\
    $m=2$  & 0.5586 & 0.5912 & 0.8731 & 0.8731 & 0.9996 & 0.9999 & 1.0003 & 1.0000 \\
    $m=3$  & 0.4887 & 0.5172 & 0.7627 & 0.7627 & 0.7241 & 0.7243 & 0.3117 & 0.3116 \\
    $m=4$  & 0.4417 & 0.4675 & 0.6524 & 0.6524 & 0.0804 & 0.0804 & 0.2414 & 0.2413 \\
    $m=5$  & 0.3990 & 0.4223 & 0.6466 & 0.6466 & 0.0673 & 0.0673 & 0.0189 & 0.0189 \\
    $m=6$  & 0.3660 & 0.3873 & 0.5543 & 0.5543 & 0.0047 & 0.0047 & 0.0167 & 0.0167 \\
    $m=7$  & 0.3342 & 0.3537 & 0.5391 & 0.5391 & 0.0037 & 0.0037 & 0.0009 & 0.0009 \\
    $m=8$  & 0.3098 & 0.3279 & 0.4949 & 0.4949 & 0.0002 & 0.0002 & 0.0008 & 0.0008 \\
    $m=9$  & 0.2905 & 0.3074 & 0.4641 & 0.4641 & 0.0002 & 0.0002 & ${\star}$ & ${\star}$ \\
    $m=10$ & 0.2747 & 0.2907 & 0.4343 & 0.4343 & ${\star}$ & ${\star}$ & ${\star}$ & ${\star}$ \\
    \hline
    \end{tabular}
    
    \caption{
        Calculated thresholds~$W_{m}$ and~$V_{m}$, including their normalized counterparts~$\zeta_{m}$ and~$\xi_{m}$, for operator families (a), (b) and (c) organized into groups of columns. Each row of the table represents a threshold obtained for states with stellar rank~$m$.
    }
    \label{res}
\end{table*}

The thresholds~\eqref{Wm-eig} and~\eqref{Vm-eig} were obtained numerically for several operator families listed in Table~\ref{refop}. The chosen families include~(a) the cubic non-linear quadrature operator, parametrized by~${\kappa}$, which is also known as the cubic state nullifier~\cite{miyata2016,brauer2021,kala2022} targeting the ideal cubic phase states. This operator represents a variance-based witness; the other three operators embody witnesses based on expectation values. The operator~(b), parametrized by~$f_{x}$ and~$f_{p}$, relates to generation of GKP states~\cite{marek2024}. Somewhat similar in purpose, (c)~targets the even and odd-parity coherent cat states~\cite{brauer2025b} with complex amplitude~$\alpha$. A simple but powerful family~(d) of Fock state nullifiers~\cite{kala2025b,kalash2025,racz2025,hotter2025} is parametrized by~$k$, the number of the target Fock state~$\ket{k}$. This is the only operator family under consideration which targets states with finite stellar rank. 

The selected operators cover both the expectation value and variance-based witnesses of non-Gaussian features, enabling a direct comparison between the two distinct approaches towards non-Gaussian resource quantification. These operators are nullifiers of the desired non-Gaussian features, embodied in specific idealized non-Gaussian states belonging to their nullspace. The hierarchy of states, determined by their stellar rank is infinite. Conversely, the stellar rank of the targeted non-Gaussian state limits the capacity of this approach, establishing an upper bound on the number of distinct thresholds recovered in~\eqref{Wm-eig} and~\eqref{Vm-eig}. Poincar\'{e} separation theorem ensures that the thresholds form \emph{non-increasing} sequences~$\{V_{m}\}_{m}$ and~$\{W_{m}\}_{m}$.

\begin{table*}[t]
    \centering
    \renewcommand{\arraystretch}{1.15}
    \setlength{\tabcolsep}{8pt}
    \begin{tabular}{|c|c|c|c|c|c|c|c|c|c|c|}

      \multicolumn{1}{c}{$\Yset_{m}$}
    & \multicolumn{1}{c}{$k=1$}
    & \multicolumn{1}{c}{$k=2$}
    & \multicolumn{1}{c}{$k=3$}
    & \multicolumn{1}{c}{$k=4$}
    & \multicolumn{1}{c}{$k=5$}
    & \multicolumn{1}{c}{$k=6$}
    & \multicolumn{1}{c}{$k=7$}
    & \multicolumn{1}{c}{$k=8$}
    & \multicolumn{1}{c}{$k=9$}
    & \multicolumn{1}{c}{$k=10$} \\
    
    \hline 
    $m=0$ & 0.611 & 1.156 & 1.612 & 2.020 & 2.396 & 2.748 & 3.081 & 3.400 & 3.705 & 4.000 \\
    $m=1$ &       & 0.545 & 0.927 & 1.239 & 1.518 & 1.776 & 2.020 & 2.251 & 2.472 & 2.685 \\
    $m=2$ &       &       & 0.516 & 0.832 & 1.083 & 1.307 & 1.515 & 1.710 & 1.896 & 2.075 \\
    $m=3$ &       &       &       & 0.500 & 0.779 & 0.997 & 1.190 & 1.368 & 1.536 & 1.695 \\
    $m=4$ &       &       &       &       & 0.490 & 0.746 & 0.942 & 1.114 & 1.273 & 1.423 \\
    $m=5$ &       &       &       &       &       & 0.483 & 0.723 & 0.904 & 1.062 & 1.207 \\
    $m=6$ &       &       &       &       &       &       & 0.477 & 0.706 & 0.875 & 1.023 \\
    $m=7$ &       &       &       &       &       &       &       & 0.473 & 0.693 & 0.854 \\
    $m=8$ &       &       &       &       &       &       &       &       & 0.470 & 0.682 \\
    $m=9$ &       &       &       &       &       &       &       &       &       & 0.467 \\
    \hline
    
    \end{tabular}
    \caption{
        Calculated non-normalized thresholds~$W_{m}$ for the Fock state family~(d). The different target Fock states~$\ket{k}$ are organized into columns. The criterion is naturally satisfied for~${m \geq k}$, with~${\zeta_{m} \equiv 0}$ whenever~${m \geq k}$. The corresponding cells are empty for visual clarity.
    }
    \label{res-fock}
\end{table*}

The thresholds~\eqref{Wm-eig} and~\eqref{Vm-eig} were calculated for states with stellar rank~${m \leq 10}$. The finite-dimensional representations of the Gaussian-transformed operators were determined analytically to avoid the unpleasant pitfalls of truncation errors~\cite{provaznik2022}, with the analytical expressions detailed within the appendix. The necessary matrices were then computed numerically on truncated Fock spaces with sufficient dimension. 

The eigenvalues of the examined matrices are finite, making it possible to upper-bound their expectation values~\cite{wolkowicz1980} and, consequently, restrict the possible values of~$\lambda$ within~\eqref{Vm-eig} to well-defined finite intervals. The minimal set of Gaussian parameters was determined for each operator family. The optimization of the parameters in both~\eqref{Wm-eig} and~\eqref{Vm-eig} was then performed by limiting their domains to reasonable ranges and dividing the space into a fine search grid. 

The results are summarized in Tables~\ref{res} and~\ref{res-fock}, where the former table covers the operator families~(a), (b), and~(c). The thresholds for cubic non-linear squeezing~(a) were obtained for~${\kappa \equiv 1}$, the thresholds for the family of GKP state nullifiers~(b) were computed with asymmetric~\cite{marek2024} parameters~$f_{p} \equiv \sqrt{\pi} \equiv 2 f_{x}$, whereas the values for the cat state nullifiers~(c) were determined for~${\alpha \equiv 2}$ in both even~($\tau_{+}$) and odd~($\tau_{-}$) parity incarnations. The parameters ($\kappa$, $f_{x}$, $f_{p}$) defining the families~(a) and~(b) can be changed arbitrarily with Gaussian transformations~\cite{kala2022,marek2024}. The particular choice of their parameters here does not incur any loss of generality. Negligible values that would be otherwise rounded down to zero are denoted with~$\star$ symbols. Finally, Table~\ref{res-fock} shows the non-normalized expectation values of the Fock state nullifier family~(d) for different target Fock state numbers~${1 \leq k \leq 10}$. The thresholds in empty cells~(${m \geq k}$) are zero as the targeted Fock states~$\ket{k}$ are naturally contained within the~$\Yset_{m \geq k}$ sets of quantum states. 

The relations for minimal values of the first two moments of the number operator attainable by Gaussian states, presented within~\cite{racz2025}, can be used to determine the thresholds on the first row by finding roots of a fourth-order polynomial.

The numerical results in both tables support the expected monotonically non-increasing trend. This behavior is in agreement with Poincar\'{e} separation theorem. As the stellar rank increases, each witness family, whether based on the expectation value or variance, attains decreasing thresholds, reflecting the growing non-Gaussian capability of the underlying state space. The thresholds for the family~(b) of GKP state nullifiers are naturally normalized by construction. The first two values are obtained analytically for infinitely squeezed vacuum states. The results show that the thresholds for families~(a), (b) and~(d) decrease at a different rate than those of~(c). The coherent cat state nullifiers quickly approach the theoretical minimal value, decreasing by several orders of magnitude across the first ranks of the infinite stellar hierarchy. This indicates that coherent cat states with~${\alpha \equiv 2}$ are, for practical purposes, indistinguishable from such approximations with relatively low stellar rank. 


These findings underline a direct quantitative connection between stellar rank and witnesses based on statistical moments of Hermitian operators. Both the normalized expectation values~$\zeta$ and normalized non-linear squeezing~$\xi$ serve as complementary and experimentally accessible tools for identifying minimal stellar rank associated with a given optical state. In practice, measuring either quantity below the corresponding normalized thresholds, either~$\zeta_m$ or~$\xi_m$, provides a clear certification that the state possesses a stellar rank at least~$m$, thus establishing a bridge between the abstract stellar hierarchy and concrete quantities based on moments of observable operators.

\section{Conclusion}

We have established a direct connection between the structural hierarchy established by stellar rank and experimentally accessible witnesses of non-Gaussian features based on expectation values and variances of observables. We demonstrated, through numerical calculations, that the normalized quantities~$\zeta_{m}$ and~$\xi_{m}$ act as reliable and complementary indicators of stellar rank, forming a consistent hierarchy of thresholds that bound the minimal stellar rank of an arbitrary single-mode optical state. 

From an experimental perspective, these quantities offer a practical route to certify non-Gaussian resources without requiring full quantum state reconstruction. Measuring a witness~$\zeta$~($\xi$) below the corresponding threshold~$\zeta_{m}$~($\xi_m$) immediately certifies that the state possesses stellar rank at least~${(m+1)}$. This approach offers a scalable and noise-tolerant method for verifying complex quantum resources using only low-order moment measurements. It opens a path toward resource-efficient quantum state certification protocols, enabling the identification of non-Gaussian resources in larger, experimentally relevant systems.

\begin{acknowledgments}
    The authors acknowledge fruitful discussions with Michal Matul\'{i}k. JP acknowledges massive use of the computational cluster of the Department of Optics and using several open-source software libraries~\cite{harris2020,virtanen2020,dalcin2021} in the computation and subsequent evaluation of the presented results.

    \emph{Funding.}
    PM, JP, and VK acknowledge the financial support of the Czech Science Foundation (project 25-17472S), and European Union’s HORIZON Research and Innovation Actions under Grant Agreement no. 101080173 (CLUSTEC).
    VK and JP acknowledges the Quantera project CLUSSTAR (8C24003) of MEYS, Czech Republic. Project CLUSSTAR has received funding from the European Union’s Horizon 2020 Research and Innovation Programme under Grant Agreements No. 731473 and No. 101017733 (QuantERA). 
    PM acknowledges a grant from the Programme Johannes Amos Comenius under the Ministry of Education, Youth and Sports of the Czech Republic reg. no. CZ.02.01.01/00/22\_008/0004649.
    SB acknowledges the support from the Carlsberg Semper Ardens project QCooL.
    VK and JF acknowledge the project IGAPrF-2025-010 of the Palacký University. VK acknowledges support from the F.R.S.–FNRS under project CHEQS within the Excellence of Science (EOS) program. 

    \emph{Data availability.}
    The complete source code, including the pre-computed datasets presented within the manuscript, is available through a GitHub~\cite{source} repository.
\end{acknowledgments}

\appendix

\section{Gaussian transformations}

\noindent The decomposition of an arbitrary Gaussian operation~$\hat{G}$ into a sequence of elementary Gaussian operations is not unique. There is a certain advantage of the particular sequence 
\begin{equation}
    \hat{G} \coloneqq \hat{F}(\vartheta) \hat{D}(z) \hat{S}(r) \hat{F}(\theta)
    \,\text{.}
\end{equation} 
The phase shift operation~${\hat{F} (\theta)}$ is passive. Its effect on the~$\hat{\Pi}_{m}$ projector is a rotation within the~${(m + 1)}$ dimensional restriction of the Fock space~\cite{fiurasek2022}. Consequently, the right-most phase shift, $F(\theta)$, can be absorbed into the eigenvalue computation in~\eqref{Wm-eig}, thus immediately reducing the number of optimization variables by one. The elementary Gaussian operations~\cite{cahill1969,kral1990} transform the bosonic annihilation and creation operators, $\hat{a}$ and~$\hat{a}^{\dagger}$ as follows,
\begin{equation}
\begin{aligned}
    \hat{F}^{\dagger}(\vartheta) \hat{a} \hat{F}(\vartheta) 
        & = \exp(\imath \vartheta) \hat{a} \\
    \hat{S}^{\dagger}(x) \hat{a} \hat{S}(r) 
        & = \cosh (r) \hat{a} - \sinh (r) \hat{a}^{\dagger} = \mu \hat{a} - \nu \hat{a}^{\dagger} \\
    \hat{D}^{\dagger}(z) \hat{a} \hat{D} (z) 
        & = \hat{a} + z 
\end{aligned}
\end{equation}
where~${\mu = \cosh(r)}$ and~${\nu = \sinh(r)}$. The order of Gaussian operations in a sequence can be exchanged~\cite{cariolaro2016}. The sequence of operations remains a Gaussian operation, only the parameters change, as summarized in Table~\ref{rel}.

\begin{table}[h]
    \def\arraystretch{1.25}
    \begin{tabular*}{\columnwidth}{@{\extracolsep{\fill}} llr}
        \textbf{Relation} & & \textbf{Reference} \\
        \hline \hline
        $\hat{D}(x) \hat{S}(r) = \hat{S}(r) \hat{D}(y)$ & $y = x \mu + \bar{x} \nu$ & \cite{cahill1969,ma1990} \\
        $\hat{D}(x) \hat{S}(r) = \hat{S}(r) \hat{D}(y)$ & $x = y \mu - \bar{y} \nu$ & \cite{cahill1969,ma1990} \\
        $\hat{D}(x) \hat{F}(\vartheta) = \hat{F}(\vartheta) \hat{D}(z)$ & $z = \emath^{-\imath \vartheta} x$ & \cite{ma1990} \\
        $\hat{F}(\vartheta) \hat{S}(r) = \hat{S}(w) \hat{F}(\vartheta)$ & $w = \emath^{2 \imath \vartheta} r$ & \cite{ma1990} \\
        $\hat{D}(x) \hat{D}(y) = \hat{D}(x + y) \exp(w)$ & $w = \frac{1}{2}(x \bar{y} - \bar{x} y)$ & \cite{cahill1969} \\
        \hline
    \end{tabular*}
    \caption{Braiding and exchange relations for Gaussian operations.}
    \label{rel}
\end{table}

\section{Gaussian Rosetta stone}

Consider a quadratic polynomial of the bosonic field operators~$\hat{Q}$ and its normal-ordered Gaussian transformation
\begin{equation}\label{GQG}
    \begin{aligned}
    \hat{Q}' 
    & = \hat{G}^{\dagger} \hat{Q} \hat{G} \\
    & = 
        A \hat{n} + 
        B \hat{a}^{2} + 
        C \hat{a}^{\dagger 2} + 
        D \hat{a} + 
        E \hat{a}^{\dagger} + 
        F
    \end{aligned}
\end{equation}
where~${\hat{n} \coloneqq \hat{a}^{\dagger}\hat{a}}$ represents the number operator. The transformed squared operator, a quartic polynomial~${\hat{G}^{\dagger} \hat{Q}^{2}\hat{G}}$, can be determined from the relation~\eqref{GQG} using Table~\ref{GQQG}. 


\section{Operator family (a)}

The non-linear quadrature operator associated with cubic non-linear squeezing, ${\hat{Q} = \hat{x} + \hat{p}^{2}}$, with canonical quadrature operators~${\sqrt{2} \hat{x} = (\hat{a} + \hat{a}^{\dagger})}$ and~${\sqrt{2} \imath \hat{p} = (\hat{a} - \hat{a}^{\dagger})}$, is a quadratic polynomial of the field operators. Its Gaussian transformation, described in terms of the factors within~\eqref{GQG}, produces
\begin{equation}
    \begin{aligned}
    A & = (g^{2} + \bar{g}^{2}) \mu \nu + \mu^{2} + \nu^{2} \\
    B & = - \frac{1}{2} (g^{2} \mu^{2} + \bar{g}^{2} \nu^{2}) - \mu \nu \\
    C & = - \frac{1}{2} (g^{2} \nu^{2} + \bar{g}^{2} \mu^{2}) - \mu \nu \\
    D & = \bar{g}^{2} \nu \bar{z} - g^{2} \mu z + \mu \bar{z} - \nu z + \frac{g \mu - \bar{g} \nu}{\sqrt{2}} \\
    E & = g^{2} \nu z - \bar{g}^{2} \mu \bar{z} + \mu z - v \bar{x} + \frac{\bar{g} \mu - g \nu}{\sqrt{2}} \\
    F & = \frac{1}{2}[ (g^{2} + \bar{g}^{2}) \mu\nu + \mu^{2} + \nu^{2} - (g^{2} z^{2} + \bar{g}^{2} \bar{z}^{2}) ] \\
      & \hphantom{=\,} + \frac{g z + \bar{g} \bar{z}}{\sqrt{2}} + \abs{z}^{2}
    \end{aligned}
\end{equation}    
where~${g = \emath^{\imath \vartheta}}$. The transformed squared operator can be determined from these factors using Table~\ref{GQQG}.

\begin{table}[t]
    \def\arraystretch{1.25}
    \begin{tabular*}{\columnwidth}{l@{\hskip 2em} l@{\extracolsep{\fill}}}
        \textbf{Element} & \textbf{Coefficient} \\
        \hline \hline
        $1$ & $F^{2} + DE + 2 BC$ \\
        $\hat{n}$ & $2 (AF + DE + BC)$ \\
        $\hat{n}^{2}$ & $A^{2} + 2 BC$ \\
        $\hat{a}$ & $AD + 2 (BE + DF)$ \\
        $\hat{a}^{2}$ & $D^{2} + 2 B (A + F)$ \\
        $\hat{a}^{3}$ & $2 BD$\\
        $\hat{a}^{4}$ & $B^{2}$ \\
        $\hat{a}^{\dagger}$ & $AE + 2 (DC + EF)$\\
        $\hat{a}^{\dagger 2}$ & $E^{2} + 2 C (A + F)$ \\
        $\hat{a}^{\dagger 3}$ & $2 CE$ \\
        $\hat{a}^{\dagger 4}$ & $C^{2}$ \\
        $\hat{n} \hat{a}$ & $2 (AD + BE)$ \\
        $\hat{n} \hat{a}^{2}$ & $2 AB$ \\
        $\hat{a}^{\dagger} \hat{n}$ & $2 (AE + CD)$ \\
        $\hat{a}^{\dagger 2} \hat{n}$ & $2 AC$ \\
        \hline
    \end{tabular*}
    \caption{
        Coefficients of the~${\hat{G}^{\dagger} \hat{Q}^{2} \hat{G}}$ operator. The individual factors~${A \dotsc F}$ refer to the expansion~\eqref{GQG} of the transformed~$\hat{G}^{\dagger} \hat{Q} \hat{G}$.
    }
    \label{GQQG}
\end{table}

\section{Operator family (b)}

The GKP state operator~\cite{marek2024} can be expressed as a linear combination of Gaussian displacement operations,
\begin{equation}
    2 - \frac{1}{2} \left[
          \{\hat{D}(\lambda_{1}) + \hat{D}^{\dagger}(\lambda_{1})\}
        + \{\hat{D}(\lambda_{2}) + \hat{D}^{\dagger}(\lambda_{2})\}
    \right]
    \,\text{,}
\end{equation}
with amplitudes~${\lambda_{1} = \sqrt{2\pi}}$ and ${\sqrt{2} \lambda_{2} = \imath \sqrt{\pi}}$ computed for the parameters~${f_{p} = \sqrt{\pi} = 2 f_{x}}$. Each displacement operator~$\hat{D}(\lambda)$ transforms according to Table~\ref{rel} into
\begin{equation}
    \hat{G}^{\dagger} \hat{D}(\lambda) \hat{G} 
     = \hat{D} (\bar{g} \mu \lambda + g \nu \bar{\lambda}) 
     \exp(\bar{g} \bar{z} \lambda - g z \bar{\lambda})
\end{equation}
where~${g = \emath^{\imath \vartheta}}$ and the exponential in the expression is only a complex number. The finite-dimensional matrix representation of the displacement operator~\cite{cahill1969} can be obtained numerically with sufficiently high accuracy.

\section{Operator family (c)}

The first component in the operator targeting coherent cat states~\cite{brauer2025b}, ${\hat{Q} = ( \hat{a}^{\dagger2} - \alpha^{*2})( \hat{a}^{2} - \alpha^{2}) + ( 1 \mp \hat{\Pi}) }$, is a quartic polynomial of the bosonic field operators. The transformation of~$\hat{n}^{2}$ is covered in the next section. The quadratic part of the polynomial, ${\hat{n} + \bar{\alpha}^{2} \hat{a}^{2} + \alpha^{2} \hat{a}^{\dagger 2}}$, transforms according to~\eqref{GQG} with factors
\begin{equation}
\begin{aligned}
    A & =   \mu^{2} + \nu^{2} - 2 \mu \nu (g^{2} \bar{\alpha}^{2} + \bar{g}^{2} \alpha^{2}) \\
    B & = - \mu \nu + \mu^{2} g^{2} \bar{\alpha}^{2} + \nu^{2} \bar{g}^{2} \alpha^{2} \\
    C & = - \mu \nu + \nu^{2} g^{2} \bar{\alpha}^{2} + \mu^{2} \bar{g}^{2} \alpha^{2} \\
    D & = - \nu z + \mu \bar{z} + 2 \mu z g^{2} \bar{\alpha}^{2} - 2 \nu \bar{z} \bar{g}^{2} \alpha^{2} \\
    E & = - \nu \bar{z} + \mu z - 2 \nu z g^{2} \bar{\alpha}^{2} + 2 \mu \bar{z} \bar{g}^{2} \alpha^{2} \\
    F & = \nu^{2} + \abs{z}^{2} - \mu \nu (g^{2} \bar{\alpha}^{2} + \bar{g}^{2} \alpha^{2}) + z^{2} g^{2} \bar{\alpha}^{2} + \bar{z}^{2} \bar{g}^{2} \alpha^{2}
\end{aligned}
\end{equation}
where~$g = \emath^{\imath \vartheta}$. The transformation of the second component can be tackled by expressing the parity operator as a phase shift~\cite{bishop1994}, $\hat{\Pi} \equiv \hat{F}(\pi)$, and employing the Gaussian braiding and exchange relations from Table~\ref{rel}. With a little algebra,
\begin{equation}
    \hat{G}^{\dagger} \hat{F}(\pi) \hat{G} = \hat{D}(y) \hat{F}(\pi)
\end{equation}
where $y = - 2 (\mu z + \nu \bar{z})$. Both the parity and the displacement operator~\cite{cahill1969} can be determined numerically with sufficiently high accuracy.

\section{Operator family (d)}

It is convenient with the expansion~${\hat{Q} = \hat{n}^{2} - 2 k \hat{n} + k^{2}}$ of the Fock state family operator~${\hat{Q} = (\hat{n} - k)^{2}}$. Because the number operator~$\hat{n}$ is invariant under phase change, the Gaussian transformation reduces to~${\hat{G} = \hat{D}(z) \hat{S}(r)}$. Its Gaussian transformation~${\hat{G}^{\dagger} \hat{n} \hat{G}}$ has the form~\eqref{GQG} with factors
\begin{equation}
    \begin{aligned}
        A & = \mu^{2} + \nu^{2}\\
        B & = - \mu\nu \\
        C & = - \mu\nu \\
        D & = \mu \bar{z} - \nu z \\
        E & = \mu z - \nu \bar{z} \\
        F & = \nu^{2} + \abs{z}^{2} \,\text{.}
    \end{aligned}
\end{equation}
The transformed~$\hat{G}^{\dagger} \hat{n}^{2} \hat{G}$ follows naturally from Table~\ref{GQQG}.

\section{The surrogate optimization problem}

Consider an operator~$\hat{Q}$ and the expression
\begin{equation}
    \begin{gathered}
        \hat{Q} - \lambda
        = \hat{Q} \pm \langle \hat{Q} \rangle - \lambda
        = \Delta \hat{Q} + [ \langle \hat{Q} \rangle - \lambda ]
    \end{gathered}
\end{equation}
where~${\Delta \hat{Q} \coloneq \hat{Q} - \langle \hat{Q} \rangle}$. Take the square of the expression,
\begin{equation}
    (\hat{Q} - \lambda)^{2}
    = \langle [\Delta \hat{Q}]^{2} \rangle + 
      [\langle \hat{Q} \rangle - \lambda]^{2} +
      2 [\Delta \hat{Q}][\langle \hat{Q} \rangle - \lambda]
\end{equation}
and its expectation value, with~$\langle\Delta \hat{Q}\rangle \equiv 0$, to retrieve
\begin{equation}
    \langle [\hat{Q} - \lambda]^{2} \rangle
    = \langle [\Delta \hat{Q}]^{2} \rangle + [\langle \hat{Q} \rangle - \lambda]^{2}
    \;\text{.}
\end{equation}
The expression is manifestly non-negative. It is convex with respect to~$\lambda$ and attains a unique global minimum at~$\lambda^{\star} \equiv \langle \hat{Q} \rangle$, equal to the variance~${\langle [\Delta\hat{Q}]^{2} \rangle}$ of the operator~$\hat{Q}$.

\vfill
\vspace*{8cm}
\bibliography{manuscript.bib}
\end{document}